\documentclass[sigconf]{acmart}

\AtBeginDocument{%
  }

\setcopyright{acmlicensed}
\copyrightyear{2018}
\acmYear{2018}
\acmDOI{XXXXXXX.XXXXXXX}

\acmISBN{978-1-4503-XXXX-X/18/06}

\begin{document}

\title{LLMTemporalComparator: A Tool for Analysing Differences in Temporal Adaptations of Large Language Models}

\author{Reinhard Friedrich Fritsch}
\email{reinhard.fritsch@student.uibk.ac.at}
\affiliation{%
  \institution{University of Innsbruck}
  \city{Innsbruck}
  \country{Austria}
}

\author{Adam Jatowt}
\email{adam.jatowt@uibk.ac.at}
\affiliation{%
  \institution{University of Innsbruck}
  \city{Innsbruck}
  \country{Austria}
}

\renewcommand{\shortauthors}{Reinhard Fritsch \& Adam Jatowt}

\begin{abstract}
This study addresses the challenges of analyzing temporal discrepancies in large language models (LLMs) trained on data from different time periods. 
To facilitate the automatic exploration of these differences, we propose a novel system that compares in a systematic way the outputs of two LLM versions based on user-defined queries. The system first generates a hierarchical topic structure rooted in a user-specified keyword, allowing for an organized comparison of topical categories. Subsequently, it evaluates the generated text by both LLMs to identify differences in vocabulary, information presentation, and underlying themes. This fully automated approach not only streamlines the identification of shifts in public opinion and cultural norms but also enhances our understanding of the adaptability and robustness of machine learning applications in response to temporal changes. By fostering research in continual model adaptation and comparative summarization, this work contributes to the development of more transparent machine learning models capable of capturing the nuances of evolving societal contexts.
\end{abstract}


\maketitle

\begin{figure*}[ht!]
    \centering
    \includegraphics[width=0.6\textwidth]{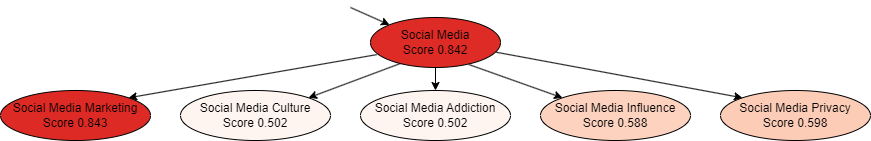}
    \caption{\centering Node Social Media is automatically divided into five subcategories 
    }
    \Description{The node Social Media is divided into five Subcategories: Marketing, Culture, Addiction, Influence and Privacy
    }
    \label{fig:subcat}
\end{figure*}

\section{Introduction}
A key focus in machine learning is the analysis and prediction of trends using large datasets. When training models on data that spans across long periods, temporal factors have a significant impact on model performance. Over time, not only do vocabulary and syntax change, but also the topics of interest, sentiments, and societal biases embedded in the data. As language evolves and social trends shift, models trained on data from different time periods may behave in noticeably different ways. These variations can provide valuable insights into how language, themes, and cultural norms change over time.

Manually identifying these temporal differences is not feasible due to the effort and time required. Our study proposes a system designed to automatically explore the temporal discrepancies between two machine learning models trained on data from different time periods\footnote{The code is available at GitHub repository: \url{https://github.com/reinhardFritsch/LLMTemporalComparator}}. The input to the tool consists of a query and the two models, which generate topical categories based on that query. The content within each category is then compared to identify differences. The aim is to contrast their outputs on the same inputs, revealing the differences that emerge due to the changing nature of data. 
By using this approach, 
we can better understand the impact of temporal shifts on machine learning applications and work towards building models that are more robust and adaptable to changes over time. Furthermore, it is possible to analyze temporal shifts in the knowledge and attention \cite{jatowt2015mapping} of continually finetuned models \cite{shi2024continuallearninglargelanguage,vandeven2019scenarioscontinuallearning} and by this to foster research in model adaptation \cite{nagabandi2019deeponlinelearningmetalearning}.

\textbf{Developing this system involves several key challenges}, such as deciding which topics to scan, selecting relevant areas for comparison, and efficiently identifying and evaluating differences, all while ensuring the system operates effectively. To address these challenges, the system is fully automated and begins with a root category defined by the user. This can be any keyword relevant to the topic of interest, such as a country, a sport, or a historical event.
%
%
The algorithm operates in two main stages:

\textbf{Category Tree Generation}: Upon initiation, the algorithm delves into the root topic by automatically constructing a hierarchical topic structure, wherein each level comprises dynamically generated subcategories. This dynamic approach, rather than relying on a predefined category tree, empowers LLMs to determine their own direction in topic exploration. This flexibility not only enhances the relevance of the generated categories but also mitigates the risk of including poorly fitting categories within the structure. These subcategories are created based on a "category chain," which is essentially a sequence of parent categories leading to the subcategory in question. This ensures that the model remains contextually aligned with the previously generated categories. A prompt template is used to direct the model, specifying the desired output format and guiding the generation of new subcategories. Each subcategory includes descriptive text, which will later be used to identify temporal differences between outputs from two models.

\textbf{Text Evaluation}: 
We use SBERT \cite{reimers-2019-sentence-bert}, which compares sentence embeddings to identify differences 
in the presented information. While this method is efficient and performs reasonably well, it may overlook more significant or interesting distinctions. 
%

We then also utilize the LLM Comparator \cite{LLM_Comparator}\footnote{https://github.com/PAIR-code/llm-comparator} as the foundation for the comparison process, making specific adaptations to fit the objectives of our study. For each comparison, the system provides both a similarity score and a detailed explanation, outlining where and why differences were identified.

\section{Approach}
For all model predictions, the Gemini Flash API \cite{gemini_flash_api} was employed. 
For evaluation, the LLM Comparator \cite{LLM_Comparator} framework was utilized. This framework was adapted to analyze each generated text via the Flash API to identify differences. 
We  
set up 
two distinct models: one  finetuned\footnote{https://github.com/ErikCikalleshi/LLM-Temporal-Difference-Learning/tree/main/Notebooks} 
on news articles from 1987 to 1995 and the other from 2000 to 2007, both based on Llama2 architecture \cite{touvron2023llama2openfoundation}\footnote{Models covering other, more recent time frames are currently being provided.}. 

\begin{figure}[h]
  \centering
  \includegraphics[width=\linewidth]{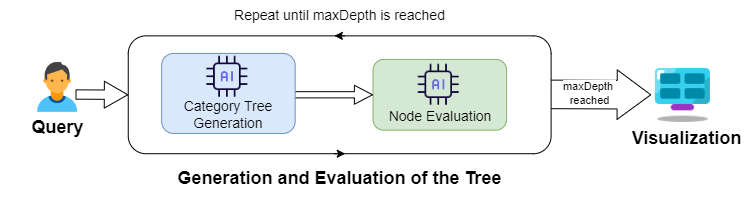}
  \caption{System Workflow}
  \label{fig:workflow}
\end{figure}

The workflow is illustrated in Figure \ref{fig:workflow} and consists of three main components: 
\begin{enumerate}
    \item Category Tree Generation
    \item Node Evaluation
    \item Visualization
\end{enumerate}

\vspace{-2em}

\subsection{Category Tree Generation} 
\begin{figure}[h]
  \centering
  \includegraphics[width=0.6\linewidth]{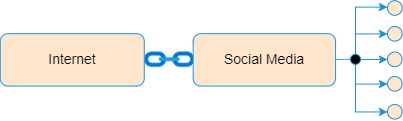}
  \caption{Category chain used to generate subcategories for root topic "Internet" and child topic "Social Media". 
  }
  \label{fig:catchain}
\end{figure}

The Category Tree Generation module begins with a root category, producing two outputs corresponding to distinct time periods of the compared models. 
Two output types are available, defined by the user at the outset: (1) generating a set of facts, which provides broader coverage, or (2) generating a paragraph, which focuses on key aspects within the category. Additionally, the number and length of subcategories can be specified. An example is shown in Figure \ref{fig:subcat}, where the parent category "Social Media" is subdivided into five subcategories. A category chain (Figure \ref{fig:catchain}) is used to maintain context during subcategory generation, ensuring each subcategory remains linked to the parent category. In this example, the model generates subcategories for the topic "Internet and Social Media." The chain for any node represents the path from the root to that node. The following prompt was used:

\textit{The input can be either a single keyword or a hierarchical chain of categories, starting from a general category and progressing to more specific ones. Your task is to generate up to {diversity} distinct categories and write a corresponding text paragraph for each. Based on the provided chain or keyword, generate up to <diversity> subcategories at the end of the chain or for the keyword, and provide an engaging and precise description for each subcategory, offering a concise summary.}

Each generation produces two outputs, consisting of the number of subcategories up to the parameter called "diversity", each with its associated text or list of facts. This is done for every node in the current layer, building a tree structure from top to bottom until the maximum depth is reached. Once a layer is completed, the algorithm revisits the generated nodes. Two cases can arise: (1) If both models generate the same category, the outputs are merged, combining both models' texts for later comparison; (2) If a category is created by one model but not the other, the second model is explicitly prompted to generate text for that category, ensuring a valid node with corresponding texts from both time periods.

\subsubsection{Node Evaluation} After each layer is generated, the node evaluation phase assigns a similarity score to each node based on the differences between the outputs. Two approaches are available for this process:
   
   \textbf{SBERT}: This method operates in two modes. In the first, SBERT \cite{reimers-2019-sentence-bert} compares the two outputs in a category, applicable to both fact and text generation. The second mode is specific to fact generation, where the algorithm matches pairs of facts from the two outputs based on similarity scores, ensuring each fact is paired with its closest match. This SBERT-based approach offers a quick overview of informational differences.
    
    \textbf{LLM Comparator}: A limitation of the SBERT approach is its lack of interpretability—while it provides similarity scores, it does not explain where or why differences occur. The LLM Comparator\footnote{https://github.com/PAIR-code/llm-comparator} addresses this by providing an interactive interface for comparing LLM-generated outputs. It evaluates text quality across several criteria and selects a "winner" based on reasoning. In our adapted framework, the focus is not on quality but on the identification of differences. Specifically, the following questions are posed to the llm-judges of the LLM Comparator framework:

    \begin{itemize}
        \item Does the responses from A and B align with data from their time frames?
        \item How big are the differences between the two inputs?
        \item How important are the differences?
        \item How interesting are the differences?
        \item Are there factual differences?
        \item Are both responses related to the question topic?
    \end{itemize}

    
    with the scoring guide shown in Table \ref{tab:score}. These labels and thresholds are inspired by the scoring table used in the original LLM-comparator framework and are automatically assigned for each generated text from different time periods. The final score is determined by averaging the scores from all evaluators and is assigned to the node. The evaluation also includes reasoning for the score and short labels summarizing the differences. An example of this will be presented in the visualization section.

\begin{table}
  \caption{Rating to score map}
  \label{tab:score}
  \begin{tabular}{ccl}
    \toprule
    label & score\\
    \midrule
    A and B are completely unrelated & 0.0 \\
    B is nonsensical or off-topic& 0.0\\
    A is nonsensical or off-topic& 0.0\\
    A and B are largely different with few similarities& 0.2 \\
    A and B share some similarities but are mostly different& 0.4\\
    A and B are somewhat similar with notable differences& 0.5\\
    A and B are fairly similar with minor differences& 0.7\\
    A and B are very similar with trivial differences& 0.9\\
    A and B are identical& 1.0\\
  \bottomrule
\end{tabular}
\end{table}

\subsection{Visualization} 
\begin{figure}
    \centering
    \includegraphics[width=.8\linewidth]{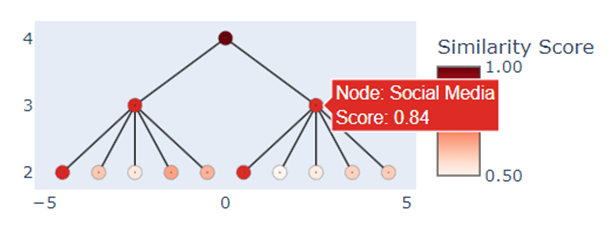}
    \caption{Tree structure with similarity score colouring.}
    \label{fig:tree}
\end{figure}

After the loop is finished and the tree created, the tree gets stored into a json file which then can be used to make two main visualizations\footnote{The examples shown in this section are minimal due to space limitation.} as follows:

\textbf{Tree structure}: A tree plot which visualizes all the nodes with colors as in Figure \ref{fig:tree}. 
 The coloring depends on either the similarity score as in the figure or as bottom up aggregation of similarity score to get a better view which category subtrees have bigger differences in them. The detailed information of a node can be viewed when clicking on the node.

\begin{figure}
    \centering
    \includegraphics[width=\linewidth]{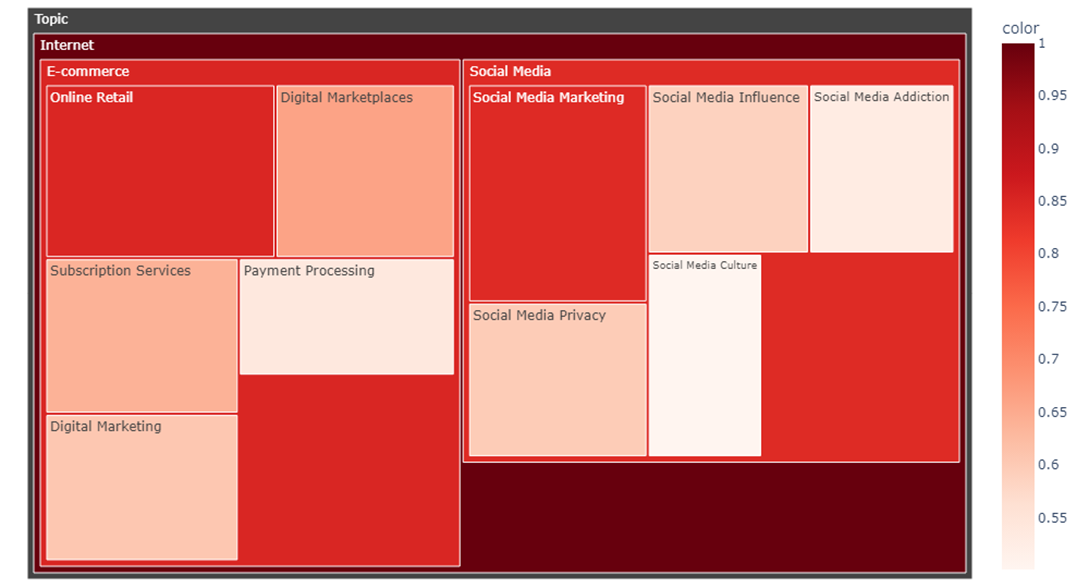}
    \caption{Treemap with similarity score colouring}
    \label{fig:treemap}
\end{figure}

\textbf{Treemap}: A treemap plot which also visualizes the structure of all categories with colors denoting similarity scores as in Figure \ref{fig:treemap}. Also here the information of each node can be viewed by clicking on it.

\begin{figure*}[ht!]
    \centering
    \includegraphics[width=.9\textwidth]{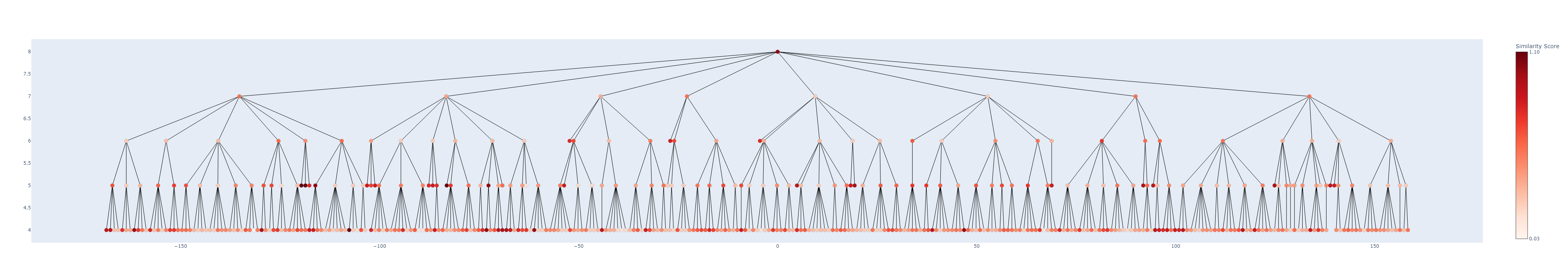}
    \caption{\centering Tree of root SeaWorld}
    \label{fig:tree:seaworld}
\end{figure*}

\subsection{Performance Improvements and Parameters} The system includes several parameters and functionalities to enhance performance, such as reducing execution time and improving result quality. These are controlled by the user via the following parameters:
\begin{itemize}
    \item \texttt{generation\_type}: Specifies either 'facts\_with\_classification' (generating multiple facts per category) or \\'paragraph\_with\_classification' (generating a single paragraph per category).
    \item \texttt{similarity\_type}: Defines the comparison way — 'facts\_with\_
    classification' compares individual facts, 'paragraph\_with\_
    classification' compares whole texts, and 'llm\_comparator' uses the LLM Comparator for reasoned comparison.
    \item \texttt{num\_samples}: Number of facts or the indicator of paragraph length per category.
    \item \texttt{tokens\_per\_fact}: Number of tokens per fact or indicator for paragraph length.
    \item \texttt{subcategory\_count}: Number of subcategories generated for each parent category. This parameter, along with depth, significantly influences runtime and the breadth of category exploration.
    \item \texttt{iteration\_depth\_for\_tree}: Defines the depth of the generated tree.
    \item \texttt{threshold\_similarity\_classes}: An important parameter (ranging from 0.0 to 1.0) used to determine when to merge two similar categories. For example, if one model generates the category 'Social Media Privacy' and the other 'Social Media Privacy's,' the system merges them to prevent duplication. This also applies when a model generates a category that is highly similar to an existing one in the tree, ensuring that redundant categories are not created.
    \item \texttt{threshold\_skip\_node\_exploration}: Another critical parameter (0.0 to 1.0) that controls when to skip subcategory generation. This helps guide the model toward areas with more differences and avoids generating further categories where similarity is already high.
\end{itemize}

An additional performance optimization could involve adjusting parameters like the previously mentioned \texttt{diversity}. 
At greater depths, categories become more specific, resulting in higher average similarity scores, and reducing diversity makes sense as the categories are already narrow in scope.

\subsection{Example}
\begin{figure}
    \centering
    \includegraphics[width=\linewidth]{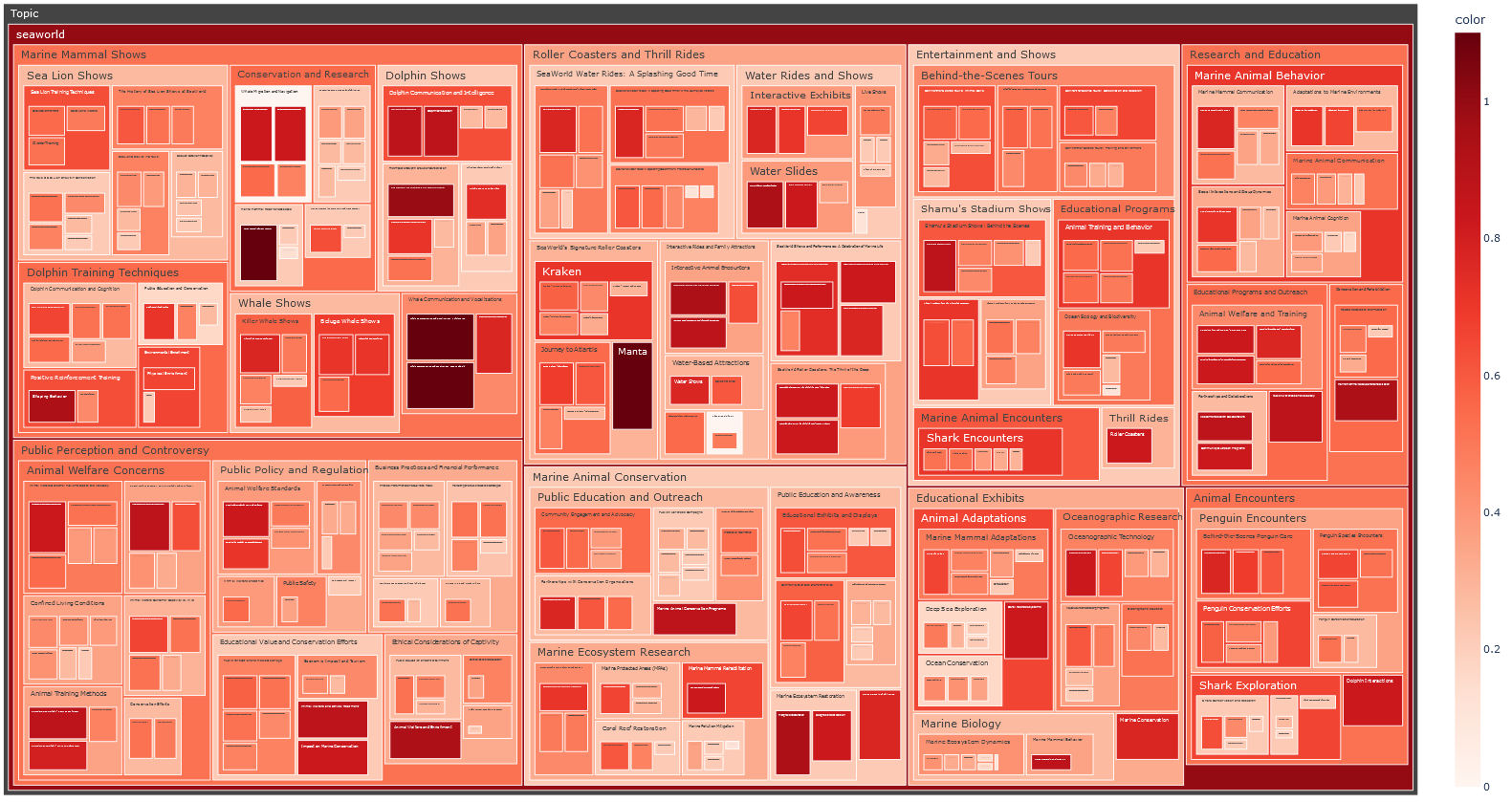}
    \caption{Treemap for query SeaWorld}
    \label{fig:treemap:seaworld}
\end{figure}

Consider a generated tree with the root topic "SeaWorld" and the more recent timeframes of 1990-1995 and 2018-2023 shown in Figure \ref{fig:tree:seaworld} as a tree and in Figure \ref{fig:treemap:seaworld} as a treemap. They have a depth of 4, containing a total of 488 categories and are of middle size with runtime of 4 hours. We will now examine the node "SeaWorld -> Educational Exhibits" along with the generated outputs.

\textbf{1990-1995}: \textit{Beyond the entertainment, SeaWorld aimed to educate visitors about the wonders of the ocean and the importance of conservation. The park featured a variety of exhibits that showcased the diverse marine life found in the world's oceans. Visitors could learn about the different species of fish, sharks, and sea turtles, as well as the threats they face from pollution, overfishing, and habitat destruction. Interactive displays, touch tanks, and behind-the-scenes tours provided a deeper understanding of marine biology and the role humans play in protecting the ocean.}

\textbf{2018-2023}: \textit{SeaWorld's educational exhibits, like the "Shamu Stadium" show, were criticized by animal rights activists in the early 2010s for their portrayal of orcas and other marine animals. The controversy led to a decline in attendance and prompted SeaWorld to phase out its orca breeding program in 2016.'}

\textbf{Example of Judge's Reasoning}: \textit{Second Response focuses on the controversy surrounding SeaWorld's orca exhibits in the early 2010s, highlighting the impact on attendance and the company's decision to phase out breeding. First Response presents a more general overview of SeaWorld's educational exhibits, focusing on the park's efforts to educate visitors about marine life and conservation. While both responses relate to SeaWorld, they present vastly different perspectives.}

In this example, four judges evaluated the similarity, with two scoring it at 0.2 and the other two at 0.4 getting a average score of 0.3. This highlights a clear shift in societal focus over time, as the controversy surrounding SeaWorld’s educational programs grew more prominent in recent years.

\section{Conclusions}
In this study, we introduced a tool to analyze and compare knowledge of language models finetuned on different over time. 
By leveraging advancements in machine learning, our approach provides researchers, policymakers, and businesses with a robust framework for understanding the dynamics of societal discourse and decision-making processes. In addition to facilitating the analysis of temporal differences, our model supports research in continual model adaptation, as highlighted in recent studies (e.g., \cite{tack2024onlineadaptationlanguagemodels}). 
Furthermore, our tool contributes to advancements in comparative summarization, enabling researchers to effectively synthesize and compare narratives across different time periods (e.g., \cite{Watanangura2023}).


Future work should focus on refining the underlying models, improving visualization techniques, and exploring additional data sources to further enhance the model’s reliability and applicability. 
\bibliographystyle{ACM-Reference-Format}
\bibliography{main}


\begin{thebibliography}{10}


\ifx \showCODEN    \undefined \def \showCODEN     #1{\unskip}     \fi
\ifx \showDOI      \undefined \def \showDOI       #1{#1}\fi
\ifx \showISBNx    \undefined \def \showISBNx     #1{\unskip}     \fi
\ifx \showISBNxiii \undefined \def \showISBNxiii  #1{\unskip}     \fi
\ifx \showISSN     \undefined \def \showISSN      #1{\unskip}     \fi
\ifx \showLCCN     \undefined \def \showLCCN      #1{\unskip}     \fi
\ifx \shownote     \undefined \def \shownote      #1{#1}          \fi
\ifx \showarticletitle \undefined \def \showarticletitle #1{#1}   \fi
\ifx \showURL      \undefined \def \showURL       {\relax}        \fi
\providecommand\bibfield[2]{#2}
\providecommand\bibinfo[2]{#2}
\providecommand\natexlab[1]{#1}
\providecommand\showeprint[2][]{arXiv:#2}

\bibitem[{Gemini Flash API}(2024)]%
        {gemini_flash_api}
\bibfield{author}{\bibinfo{person}{{Gemini Flash API}}.} \bibinfo{year}{2024}\natexlab{}.
\newblock \bibinfo{title}{API Documentation}.
\newblock \bibinfo{howpublished}{\url{https://ai.google.dev/gemini-api/docs?hl=de}}.
\newblock
\newblock
\shownote{Accessed: 2024-09-25}.


\bibitem[Jatowt et~al\mbox{.}(2015)]%
        {jatowt2015mapping}
\bibfield{author}{\bibinfo{person}{Adam Jatowt}, \bibinfo{person}{{\'E}milien Antoine}, \bibinfo{person}{Yukiko Kawai}, {and} \bibinfo{person}{Toyokazu Akiyama}.} \bibinfo{year}{2015}\natexlab{}.
\newblock \showarticletitle{Mapping temporal horizons: Analysis of collective future and past related attention in Twitter}. In \bibinfo{booktitle}{\emph{Proceedings of the 24th international conference on World Wide Web}}. \bibinfo{pages}{484--494}.
\newblock


\bibitem[Kahng and et~al.(2024)]%
        {LLM_Comparator}
\bibfield{author}{\bibinfo{person}{Minsuk Kahng} {and} \bibinfo{person}{et al.}} \bibinfo{year}{2024}\natexlab{}.
\newblock \showarticletitle{LLM Comparator: Visual Analytics for Side-by-Side Evaluation of Large Language Models}. In \bibinfo{booktitle}{\emph{Extended Abstracts of the 2024 CHI Conference on Human Factors in Computing Systems}} \emph{(\bibinfo{series}{CHI EA '24})}. \bibinfo{publisher}{Association for Computing Machinery}, \bibinfo{address}{New York, NY, USA}, Article \bibinfo{articleno}{216}, \bibinfo{numpages}{7}~pages.
\newblock
\showISBNx{9798400703317}
\urldef\tempurl%
\url{https://doi.org/10.1145/3613905.3650755}
\showDOI{\tempurl}


\bibitem[Nagabandi et~al\mbox{.}(2019)]%
        {nagabandi2019deeponlinelearningmetalearning}
\bibfield{author}{\bibinfo{person}{Anusha Nagabandi}, \bibinfo{person}{Chelsea Finn}, {and} \bibinfo{person}{Sergey Levine}.} \bibinfo{year}{2019}\natexlab{}.
\newblock \bibinfo{title}{Deep Online Learning via Meta-Learning: Continual Adaptation for Model-Based RL}.
\newblock
\newblock
\showeprint[arxiv]{1812.07671}~[cs.LG]
\urldef\tempurl%
\url{https://arxiv.org/abs/1812.07671}
\showURL{%
\tempurl}


\bibitem[Reimers and Gurevych(2019)]%
        {reimers-2019-sentence-bert}
\bibfield{author}{\bibinfo{person}{Nils Reimers} {and} \bibinfo{person}{Iryna Gurevych}.} \bibinfo{year}{2019}\natexlab{}.
\newblock \showarticletitle{Sentence-BERT: Sentence Embeddings using Siamese BERT-Networks}. In \bibinfo{booktitle}{\emph{Proceedings of the 2019 Conference on Empirical Methods in Natural Language Processing}}. \bibinfo{publisher}{Association for Computational Linguistics}.
\newblock
\urldef\tempurl%
\url{https://arxiv.org/abs/1908.10084}
\showURL{%
\tempurl}


\bibitem[Shi et~al\mbox{.}(2024)]%
        {shi2024continuallearninglargelanguage}
\bibfield{author}{\bibinfo{person}{Haizhou Shi}, \bibinfo{person}{Zihao Xu}, \bibinfo{person}{Hengyi Wang}, \bibinfo{person}{Weiyi Qin}, \bibinfo{person}{Wenyuan Wang}, \bibinfo{person}{Yibin Wang}, \bibinfo{person}{Zifeng Wang}, \bibinfo{person}{Sayna Ebrahimi}, {and} \bibinfo{person}{Hao Wang}.} \bibinfo{year}{2024}\natexlab{}.
\newblock \bibinfo{title}{Continual Learning of Large Language Models: A Comprehensive Survey}.
\newblock
\newblock
\showeprint[arxiv]{2404.16789}~[cs.LG]
\urldef\tempurl%
\url{https://arxiv.org/abs/2404.16789}
\showURL{%
\tempurl}


\bibitem[Tack et~al\mbox{.}(2024)]%
        {tack2024onlineadaptationlanguagemodels}
\bibfield{author}{\bibinfo{person}{Jihoon Tack}, \bibinfo{person}{Jaehyung Kim}, \bibinfo{person}{Eric Mitchell}, \bibinfo{person}{Jinwoo Shin}, \bibinfo{person}{Yee~Whye Teh}, {and} \bibinfo{person}{Jonathan~Richard Schwarz}.} \bibinfo{year}{2024}\natexlab{}.
\newblock \bibinfo{title}{Online Adaptation of Language Models with a Memory of Amortized Contexts}.
\newblock
\newblock
\showeprint[arxiv]{2403.04317}~[cs.LG]
\urldef\tempurl%
\url{https://arxiv.org/abs/2403.04317}
\showURL{%
\tempurl}


\bibitem[Touvron and et~al.(2023)]%
        {touvron2023llama2openfoundation}
\bibfield{author}{\bibinfo{person}{Hugo Touvron} {and} \bibinfo{person}{et al.}} \bibinfo{year}{2023}\natexlab{}.
\newblock \bibinfo{title}{Llama 2: Open Foundation and Fine-Tuned Chat Models}.
\newblock
\newblock
\showeprint[arxiv]{2307.09288}~[cs.CL]
\urldef\tempurl%
\url{https://arxiv.org/abs/2307.09288}
\showURL{%
\tempurl}


\bibitem[van~de Ven and Tolias(2019)]%
        {vandeven2019scenarioscontinuallearning}
\bibfield{author}{\bibinfo{person}{Gido~M. van~de Ven} {and} \bibinfo{person}{Andreas~S. Tolias}.} \bibinfo{year}{2019}\natexlab{}.
\newblock \bibinfo{title}{Three scenarios for continual learning}.
\newblock
\newblock
\showeprint[arxiv]{1904.07734}~[cs.LG]
\urldef\tempurl%
\url{https://arxiv.org/abs/1904.07734}
\showURL{%
\tempurl}


\bibitem[Watanangura et~al\mbox{.}(2023)]%
        {Watanangura2023}
\bibfield{author}{\bibinfo{person}{Patcharapruek Watanangura}, \bibinfo{person}{Sukit Vanichrudee}, \bibinfo{person}{On Minteer}, \bibinfo{person}{Theeranat Sringamdee}, \bibinfo{person}{Nattapong Thanngam}, {and} \bibinfo{person}{Thitirat Siriborvornratanakul}.} \bibinfo{year}{2023}\natexlab{}.
\newblock \showarticletitle{A Comparative Survey of Text Summarization Techniques}.
\newblock \bibinfo{journal}{\emph{SN Computer Science}} \bibinfo{volume}{5}, \bibinfo{number}{1} (\bibinfo{date}{02 Dec} \bibinfo{year}{2023}), \bibinfo{pages}{47}.
\newblock
\showISSN{2661-8907}
\urldef\tempurl%
\url{https://doi.org/10.1007/s42979-023-02343-6}
\showDOI{\tempurl}


\end{thebibliography}

\appendix

\end{document}